\documentclass[conference]{IEEEtran}
\usepackage{cite}

\ifCLASSINFOpdf
\usepackage[pdftex]{graphicx}
\else
\fi
\usepackage[cmex10]{amsmath}
\usepackage{fixltx2e}
\usepackage{url}

\newtheorem{thm}{Theorem}
\newtheorem{definition}[thm]{Definition}

\usepackage{amssymb}

\let\emptyset\varnothing

\newcommand{\bhvr}[1]{\hbox{${\beta_{\scriptsize #1}}$}}
\newcommand{\bhvrtwo}[2]{\hbox{${\beta^{\scriptsize #2}_{\scriptsize #1}}$}}

\usepackage{pbox}

\begin{document}
%
\title{On the Behavioral Interpretation of System-Environment Fit and Auto-Resilience}

\author{\IEEEauthorblockN{Vincenzo De Florio}
\IEEEauthorblockA{PATS research group,
University of Antwerp \& iMinds Research Institute\\
Middelheimlaan 1, 2020 Antwerpen, Belgium\\
Email: https://www.uantwerp.be/en/staff/vincenzo-deflorio}
}


%


\maketitle

\begin{abstract}
Already 71 years ago Rosenblueth, Wiener, and Bigelow
introduced the concept of
the ``behavioristic study of natural events'' and proposed
a classification of systems according to the quality
of the behaviors they are able to exercise.
In this paper we consider the problem of the resilience
of a system when deployed in a changing environment,
which we tackle by considering the behaviors both
the system organs and the environment mutually exercise.
We then introduce a partial order and a metric space
for those behaviors, and we use them to define
a behavioral interpretation of the concept of 
system-environment fit.
Moreover we suggest that behaviors based on the extrapolation
of future environmental requirements would allow systems to proactively
improve their own system-environment fit and optimally evolve
their resilience.
Finally we describe how we plan to express a complex optimization
strategy in terms of the concepts introduced in this paper.
\end{abstract}



%
\IEEEpeerreviewmaketitle

\section{Introduction}\label{s:intro}

Let us consider a familiar case of ``systems'': the human beings. Human beings are generally considered as the highest 
peak of biological evolution. Their behavioral and teleological characteristics~\cite{RWB43} set them apart from 
other system classes~\cite{Bou56} and make them appear to be more ``gifted'' than other beings, e.g., dogs. But how 
do the superior qualities of mankind translate in terms of \emph{resilience}? Under stressful or turbulent 
conditions we know that 
often a man will result ``better'' than a dog: superior awareness, consciousness, manual and technical dexterity, 
and reasoning; advanced ability to reuse experience, learn, develop science, as well as other factors, they all 
lead to the apparently ``obvious'' conclusion that mankind has a greater ability to tolerate adverse 
conditions.

And though, it is also quite easy to find counterexamples. If a threat, e.g., comes with ultrasonic noise, a dog 
may perceive the threat and react---for instance by running away---while a man may stay unaware until too late. 
Or consider the case of miners: inability to perceive toxic gases makes them vulnerable to, e.g., carbon monoxide 
and dioxide, methane, and other lethal gases~\cite{DF13b}. A simpler system able to perceive the threat and flee would 
have more chances to survive. Perception of course is but one of a number of ``systemic features'' that need to be
available in order to counterbalance a threat.

So how do we tell whether a system is fit to stand the new conditions
characterizing a changing environment?
How do we reason about the quality of resilience? And, even more importantly, how
do we make sure that a system ``stays fit'' if the environment changes?

The above questions are discussed and, to some extent, addressed in this paper.

Our starting point here is the conjecture that
	\emph{resilience is no absolute figure}; rather, it is \emph{the result of a match 
with a deployment environment}.
Whatever its structure, organization, architecture, capabilities, and resources, a system is only robust as long 
as its ``provisions'' (its system characteristics, including the ability to develop knowledge and ``wisdom'') 
match the current environmental conditions.

A second cornerstone of the present discussion is given by the assumption that the interactions between
systems and environments can be expressed and reasoned upon by considering the behaviors expressed during
those interactions.
In other words, a system-environment fit is the result of the match between the behaviors exercised by a system
and those exercised by its environment (including other systems, the users, etc.)

A third and final assumption is that reasoning about a system's resilience is facilitated by
considering the behaviors of those system ``organs'' (namely, sub-systems) responsible for the
following abilities:


\begin{enumerate}
\item the ability to perceive change;
\item the ability to ascertain the consequences of change; 
\item the ability to plan a line of defense against threats deriving from change;
\item the ability to enact the defense plan being conceived in step 3;
\item and, finally, the ability to treasure up past experience and continuously improve, to some extent, abilities 1--4.
\end{enumerate}

As can be clearly seen, the above abilities correspond to the components of the so-called MAPE-K loop of autonomic
computing~\cite{KeCh:2003}.
We shall refer to those abilities as well as the organs that embed them as to
the ``systemic features.''

In what follows we first focus in Sect.~\ref{s:sf} on the concept of behavior
and recall the five major classes of behaviors according to
Rosenblueth, Wiener, and Bigelow~\cite{RWB43} and Boulding~\cite{Bou56}.
We then introduce a system's \emph{cybernetic class\/} by associating
each of the systemic features with its own behavior class.

After this, in Sect.~\ref{s:fit}, we introduce a behavioral formulation of the concepts
of supply and system-environment fit
as measures of the optimality of a given
design with respect to the current environmental conditions.

Section~\ref{s:or} then suggests how proactive and/or social behaviors that
would be able to track supply and system-environment fit would pave the way
to systems able to self-tune their systemic features in function of the experienced
or predicted environmental conditions.

An application of the concepts presented in this work is briefly described in Sect.~\ref{s:ls}.

Our conclusions are finally stated in Sect.~\ref{s:end}.

\section{Systemic Features}\label{s:sf}
As mentioned above, an important attribute towards achieving robustness is given by what we called 
in Sect.~\ref{s:intro} as
the ``systemic features'', or the behaviors typical of the system under scrutiny. Such behaviors are
the subject of the present section.

In what follows we first recall in Sect.~\ref{s:sf:bc} what are the main behavioral classes.
The main sources here are the classic works by Rosenblueth, Wiener, and Bigelow~\cite{RWB43} and Boulding~\cite{Bou56}.
In the first work, classes were
identified by the Authors by considering the system in isolation. In the second one
Boulding introduced an additional class considering the social dimension.

After this, in Sect.~\ref{s:sf:cc}, we consider an exemplary system; we identify
in it the main system organs responsible for resilience; and associate behavioral classes to those organs.
By doing so we characterize $C$, namely the ``cybernetic class'' of the system under consideration.

\subsection{Behavioral Classes}\label{s:sf:bc}
Already 71 years ago Rosenblueth, Wiener, and Bigelow~\cite{RWB43} introduced the concept of
the ``behavioristic study of natural events'', namely ``the examination of the
output of the object and of the relations of this output to the input''. The term ``object''
in the cited paper corresponds to that of ``system''.
In that renowned text the Authors purposely 
``omit the specific structure and the intrinsic organization'' of the systems under scrutiny
and classify them exclusively on the basis of the quality of the ``change produced
in the surroundings by the object'', namely the system's behavior.
The Authors identify in particular four major classes of behaviors\footnote{For the sake
	of brevity we will not discuss here passive behavior.}:
\begin{description}
	\item[\bhvr{\hbox{ran}}\,:] Random behavior. This is an active form of behavior that does not appear to serve
		a specific purpose or reach a specific state. A source of electro-magnetic interference 
		exercises random behavior.
	\item[\bhvr{\hbox{pur}}\,:]
		Purposeful behavior. This is behavior that serves a purpose and is directed towards
		a specific goal. Quoting the Authors, in purposeful behavior we can observe a
		``final condition toward which the movement [of the object] strives''. Servo-mechanisms are examples of 
		purposeful behavior.
	\item[\bhvr{\hbox{rea}}\,:]
		Reactive behavior. This is behavior that
		``involve[s] a continuous feed-back from the goal that modifies and
		guides the behaving object''. 
		Examples of this behavior include phototropism, namely the tendency we observe, e.g., in certain plants,
		to grow towards the light, and gravitropism, viz. the tendency of plant roots to grow downward.
		Reactive behaviors require the system to be open~\cite{Hey98} (able that is to
		continuously perceive, communicate, and interact with external systems and the
		environment) and to embody some form of feedback loop.
	\item[\bhvr{\hbox{pro}}\,:]
		Proactive behavior. This is behavior directed towards the extrapolated future state of the goal.
		The Authors in~\cite{RWB43} classify proactive behavior according to its ``order'', namely
		the amount of context variables taken into account in the extrapolation.
\end{description}

Kenneth Boulding in his classic paper~\cite{Bou56} introduces an additional class:
\begin{description}
	\item[\bhvr{\hbox{soc}}\,:]
		Social behaviors.
		This class is based on the concept of social organization.
		Quoting the Author, in such systems ``the unit is not perhaps the person---the individual human as such---but the `role'---that part of the person which is concerned with the organization or situation in question, and it is tempting to define social organizations, or almost any social system, as a set of role tied together with channels of communication.''
		Social behaviors may take different forms and be, e.g., mutualistic, commensalistic, co-evolutive, 
		or co-opetitive~\cite{BN98,AK10,AF83}. For more information
		we refer the Reader to~\cite{2014arXiv1401.5607D}.
\end{description}

We shall define $\pi$ as a projection map returning, for each of the above behavior classes, 
an integer in $\{1,\ldots,5\}$ ($\pi(\bhvr{\hbox{ran}})=1$, \ldots, $\pi(\bhvr{\hbox{soc}})=5$).

For any behavior \bhvr{x}{} and any set of context figures $F$, notation \bhvrtwo{x}{F}{} will be used to denote that \bhvr{x}{} is exercised
by considering the context figures in $F$. Thus if, for instance, $F=(\hbox{speed},\hbox{luminosity})$, then
\bhvrtwo{\hbox{rea}}{F}{} refers to a reactive behavior that responds to changes in speed and light.

For any behavior \bhvr{x}{} and any integer $n>0$, notation \bhvrtwo{x}{n}{} will be used to denote that \bhvr{x}{} is exercised
by considering $n$ context figures, without specifying which ones.

As an example,
behaviour \bhvrtwo{\hbox{pro}}{|F|}{}, with $F$ defined as above,
identifies an 
order\mbox{-}\nobreak\hspace{0pt}2
proactive behavior while \bhvrtwo{\hbox{pro}}{F}{} says in addition that that
behavior considers both speed and luminosity 
to extrapolate the future position of the goal.

We now introduce the concept of partial order among behaviors.

\begin{definition}[Partial order of behaviors]\label{d:partial.order}
Given any two behaviors $\beta_1$ and $\beta_2$ we shall say that $\beta_1 \prec \beta_2$ if and only if
either of the following conditions holds:
  \begin{enumerate}
  \item $\pi(\beta_1) < \pi(\beta_2)$.
  \item $\left(\pi(\beta_1) = \pi(\beta_2)\right) \wedge$\\
	  $\left(\exists (F, G): \beta_1=\bhvrtwo{1}{F} \wedge \beta_2=\bhvrtwo{2}{G} \wedge F \subsetneq G\right)$.
  \item $\left(\pi(\beta_1) = \pi(\beta_2) = \bhvr{\hbox{pro}}\right) \wedge$\\
	  $\left(\exists (n,m): \beta_1=\bhvrtwo{1}{n} \wedge \beta_2=\bhvrtwo{2}{m} \wedge n<m\right)$.
  \end{enumerate}
\end{definition}

\typeout{In other words. . .}

Whenever two behaviors $\beta_1$ and $\beta_2$ are such that $\beta_1 \prec \beta_2$, it is possible to
define some notion of distance between the two behaviors by considering an arithmetization\footnote{A classic
	example of arithmetization may be found in the renowned work~\cite{Goe31} by Kurt G\"odel.}
based on, e.g., the following factors used as exponents of three different prime numbers:
  \begin{enumerate}
  \item $\pi(\beta_2) - \pi(\beta_1)$.
  \item $|G\setminus F|$.
  \item $m-n$.
  \end{enumerate}

In what follows we shall assume that some metric function, $\mathbf{dist}$, has been defined.

\subsection{Cybernetic Class}\label{s:sf:cc}
The behavioral classes recalled in~\ref{s:sf:bc} may be applied to the five ``systemic features'' introduced
in Sect.~\ref{s:intro}.
For any system $s$ we shall refer to the systemic features of $s$ through the following 5-tuple:
\begin{equation}
	\left(C_M(s), C_A(s), C_P(s), C_E(s), C_K(s)\right),\label{e:sf}
\end{equation}
whose components orderly correspond to the abilities introduced in Sect.~\ref{s:intro} as well as
to the stages of MAPE-K loops~\cite{KeCh:2003}.
System $s$ will me omitted when it can be implicitly identified without introducing ambiguity.

\begin{definition}[Cybernetic Class]
For any given system $s$
we define as cybernetic class the 5-tuple
\begin{equation}
	C(s) = \left(\beta_{C_M(s)}, \beta_{C_A(s)}, \beta_{C_P(s)}, \beta_{C_E(s)}, \beta_{C_K(s)}\right),\label{e:cc}
\end{equation}
where, for any $x\in\{M,A,P,E,K\}$, $\beta_{C_x(s)}$ represents the behavior class assigned to
systemic feature $C_x$ of $s$,
or $\emptyset$ if $s$ does not include $C_x$ altogether.
\end{definition}

As can be clearly understood,
a system's cybernetic class is a qualitative metric that does not provide a full coverage of the systemic characteristics of the system.
As such it should be complemented with quantitative assessments of the quality of service of
its system organs---namely the sub-systems responsible for hosting
its systemic features \eqref{e:sf}. 
In particular for $C_M(s)$ and $C_E(s)$---namely, the features corresponding to the abilities
of perception and actuation---it is useful to complement the notion of behavior with a characterization of the set of context variables
that are under the ``sphere of action'' of the corresponding organs. For $C_M(s)$ this means specifying the set of
context figures that may be timely perceived by $s$~\cite{DF12a,DF13b}. Interestingly enough, this concept
closely corresponds to that of the powers of representation in Leibniz~\cite{leibniz2006shorter}.
When considering $C_E(s)$, the sphere of action could be represented by the set of the context figures that may be
controlled---to a certain extent---through system behaviors.

We observe that features $C_M$ and $C_E$ are intrinsically purposeful.
We believe that notation \bhvrtwo{\hbox{pur}}{F}{} provides a convenient and homogeneous way to
express the behavior class and the spheres of action of both $M$ and $E$ organs.

It is now possible to characterize a system's cybernetic class through notation~\eqref{e:cc}. As an example,
by following the assessments proposed in~\cite{DF13a},
the adaptively redundant data structures described in~\cite{DB07a} have the following cybernetic class
\[ C_1 = (\bhvr{\hbox{pur}}, \bhvrtwo{\hbox{pro}}{1}, \bhvr{\hbox{pur}}, \bhvr{\hbox{pur}}, \emptyset), \]
while the adaptive $N$-version programming system introduced in~\cite{BDB11a,Buys2012c} is
\[ C_2 = (\bhvr{\hbox{pur}}, \bhvrtwo{\hbox{pro}}{2}, \bhvr{\hbox{pur}}, \bhvr{\hbox{pur}}, \bhvr{\hbox{pur}}). \]

We believe the notion and notation of cybernetic class provide a
convenient way to compare qualitatively the systemic features of any two systems
with reference to their robustness. As an example,
by comparing the above 5-tuples $C_1$ and $C_2$ one may easily realize how the major strength
of those two systems lies in their analytic organs, both of which are capable of proactive behaviors
(\bhvr{\hbox{pro}})---though in a simpler fashion in $C_1$. Another noteworthy difference is the presence
of a knowledge organ in $C_2$, which indicates that the second system is able to accrue and make use
of the past experience in order to improve its action---to some extent and exclusively through \bhvr{\hbox{pur}}{} behaviors.
We conjecture that the action of the knowledge organ in this case corresponds to so-called
\emph{antifragility\/}~\cite{Taleb12,DF14a}, namely the
ability to ``treasure up'' the past experience so as to improve one's system-environment fit.

\section{System-Environment Fit}\label{s:fit}
What presented in Sect.~\ref{s:sf} allows for a system to be characterized---to some extent---in terms of its
``systemic features''---the provisions that is that play a role when responding to change.
As a way to identify the ``quality'' of those provisions in that section we made use of the different behavioral classes as
defined in~\cite{RWB43,Bou56}, and introduced $C(s)$ as well as its
components.

Here we move our attention to a second aspect that, we conjecture, needs to be
considered when assessing a system's resilience. This second aspect tells us how the 
cybernetic class
matches the requirements of dynamically changing environmental conditions.

As already anticipated in Sect.~\ref{s:intro}, in what follows we assume that the evolution
of an environment may also be expressed as
a behavior. Said behavior may be of any of the types listed in Sect.~\ref{s:sf:bc} and as such
it may result in the dynamic variation of a number of ``firing context figures''. In fact those figures characterize
and, in a sense, set the boundaries of an \emph{ecoregion}, namely
``an area defined by its environmental conditions''~\cite{Dictionary.com2014}.

An environment may be the result of the action of, e.g., 
a human being (a ``user''), or a software managing an ambient, or for instance
it may be the result of purposeless (random) behavior---such as a source of electro-magnetic interference.
As a consequence, an environment may behave randomly or exhibit a recognizable trend; in the latter case the
variation of its context figures may be such that it allows for tracking or speculation
(extrapolation of future states).
Moreover, an environment may exhibit the same
behavior for a relatively long period of time or it may vary dynamically its character.

We shall refer in what follows to the dynamic evolution of
environmental behavior as to an environment's \textbf{turbulence}.

\begin{figure}
	\includegraphics[width=0.5\textwidth]{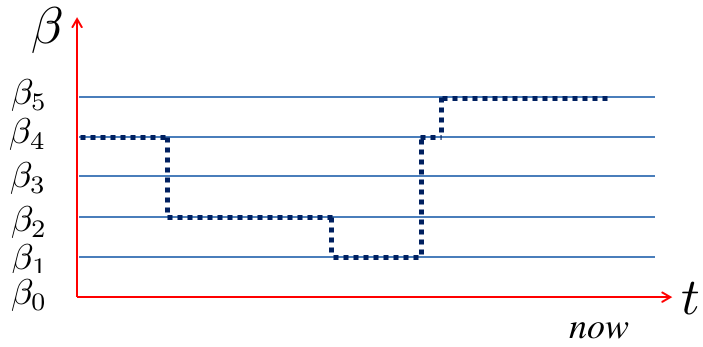}
	\caption{Exemplification of turbulence, namely the dynamic evolution of environmental behavior (shown here
	as a
	dotted line). Abscissas are time, ``\emph{now}'' being
	the current time. Ordinates are the behavior classes exercised by the environment.}\label{f:env}
\end{figure}

Diagrams such as 
the one in Fig.~\ref{f:env} may be used
to represent the dynamic evolution of environments.

%


%
%

It is now possible to propose a definition of two indicators for the
quality of resilience: the system supply relative to an environment
and the system-environment fit.

\begin{definition}[System supply]\label{d:supply}
	Given a system $S$ deployed in an environment $E$, characterized respectively by
	behaviors $\beta^S(t)$ and $\beta^E(t)$;
	and given a metric function $\mathbf{dist}$;
	we define as supply at time $t$ with respect to $\beta^E(t)$
	the following value:
	\begin{multline}
	\mathbf{supply}(S,E,t) = \nonumber \\
	= \begin{cases} 
		 \mathbf{dist}(\beta^S(t),\beta^E(t))  & \mbox{if } \beta^E(t)\prec \beta^S(t)\\
		-\mathbf{dist}(\beta^S(t),\beta^E(t))  & \mbox{if } \beta^S(t)\prec \beta^E(t)\\
		0 & \pbox{0.5\textwidth}{\mbox{if  $\beta^E(t)$ and $\beta^S(t)$ ex-}\\
					 \mbox{press the same behaviors.}}
	\end{cases}
	\end{multline}
\end{definition}

Supply can be positive (oversupply), negative (undersupply), or zero (perfect supply).

\begin{definition}[System-environment fit]\label{d:fit}
	Given the same conditions as in Definition~\ref{d:supply},
	we define as the system-environment fit at time $t$ the function
	\begin{multline}
		\mathbf{fit}(S,E,t) = \\ \nonumber \\
	= \begin{cases} 
		1 / (1 + \mathbf{supply}(S,E,t) ) & \mbox{if } \mathbf{supply}(S,E,t) \ge 0\\
		-\infty & \mbox{otherwise.}\\
	\end{cases}
	\end{multline}
\end{definition}

The above definition expresses system-environment fit as a function returning 1 in the case
of best fit; slowly scaling down with oversupply; and returning $-\infty$ in case of undersupply.
It is not the only possible such definition of course: an alternative one is given, for instance,
by having $\mathbf{supply}^2$ instead of $\mathbf{supply}$.

\begin{figure}
	\includegraphics[width=0.5\textwidth]{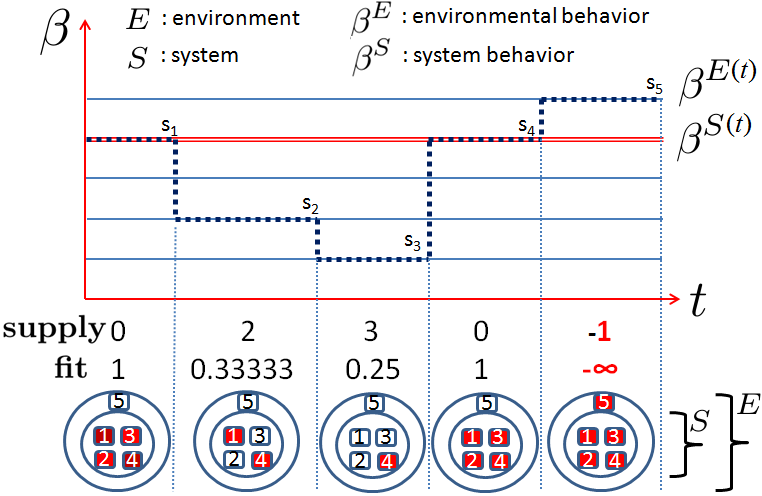}
	\caption{Exemplification of supply and system-environment fit.}
	\label{f:fitset}
\end{figure}

Figure~\ref{f:fitset} exemplifies a system-environment fit in the case of two behaviors $\beta^S$ and $\beta^E$
with $S\subsetneq E$. $E$ consists of five context figures identified by integers $1,\dots,5$ while $S$
consists of context figures $1,\dots,4$.  The system behavior is assumed to be constant;
if $S=C(M)$ this means that the system's perception organ constantly monitors the four figures $1,\dots,4$.
On the contrary $\beta^E$ varies with time. Five time segments are exemplified ($s_1,\dots,s_5$) during
which the following context figures are affected:
\begin{description}
	\item[$s_1$]: Figures $1,\dots,4$.
	\item[$s_2$]: Figure $1$ and figure $4$.
	\item[$s_3$]: Figure $4$.
	\item[$s_4$]: Figures $1,\dots,4$.
	\item[$s_5$]: Figures $1,\dots,5$.
\end{description}
Figures are represented as boxed integers, with an empty box meaning that the figure is not affected by the
environment and a filled box
meaning the figure is affected. The behaviour of the environment is constant within a time segment and changes
at the next one.
This is shown through the sets at the bottom of Fig.~\ref{f:fitset}:
for each segment $t_s\in \{s_1,\dots,s_5\}$ the superset is $E(t_s)$ while the subset is $S(s_t)$, namely $E(s_t)\cap S$.
The relative supply and the system-environment fit also change with the time segments.
During $s_1$ and $s_4$ there is perfect supply and best fit: the behavior exercised by the environment is evenly
matched by the features of the system. During $s_2$ and $s_3$ the systemic features are more than enough
to match the current environmental conditions---a case of what we referred to as ``oversupply''. Correspondingly,
fit is rather low.
In $s_5$ we have the opposite situation: the systemic features---for instance, pertaining to a perception organ---are
insufficient to become aware of all the changes produced by the environment. In particular here
changes connected with figure 5 go undetected. This is a case of ``undersupply'', corresponding
to the ``worst possible'' system-environment fit.

\section{Optimal Resilience}\label{s:or}
The two functions introduced in Sect.~\ref{s:fit}, $\mathbf{supply}$ and $\mathbf{fit}$, may be
interpreted as measures of the optimality of a given design with respect to the current
environmental conditions. Whenever those conditions allow it and a partial order ``$\prec$''
exists for the behaviors at play, then it is possible to consider system behaviors of the following forms:

\begin{enumerate}
	\item
	$\bhvrtwo{\hbox{pro}}{F}$, with $F$ including figures $\mathbf{supply}$ and $\mathbf{fit}$.\label{e:opt.pro}

Such behavior,
when exercised by system organs for analysis, planning, and knowledge management,
translates in the possibility to become aware and speculate on the possible future robustness
requirements. If this is coupled with the possibility to revise one's system organs by
enabling or disabling, e.g., the ability to perceive certain context figures depending
on the extrapolated future environmental conditions, then a system could proactively
improve its own system-environment fit.

\item
	$\bhvrtwo{\hbox{soc}}{F}$, with $F$ including figures $\mathbf{supply}$ and $\mathbf{fit}$.\label{e:opt.soc}

Analysis, planning, and knowledge management behaviors of this type aim at
artificially augmenting or reducing
the system features by establishing / disestablishing collaborative relationships as exemplified
in the ``canary-in-the-coal-mine'' scenario of~\cite{DF13b}.
\end{enumerate}

As we did in the paper just cited we propose to call behaviors such as~\ref{e:opt.pro}) and~\ref{e:opt.soc})
as \textbf{auto-resilient}.

Finally, we remark how the formulation of system-environment fit presented in this work
may also be tailored so as to include overheads and costs.

\section{An Application: Project LittleSister}\label{s:ls}
LittleSister~\cite{LS} is an ICON project financed by the iMinds research institute and the Flemish 
Government Agency for Innovation by Science and Technology (IWT).  
The project aims to deliver a low-cost telemonitoring~\cite{Meystre05} solution for home 
care and is to run until the end of year 2014.  
LittleSister adopts a connectionist approach
in which the collective action of an interconnected network
of simple units~\cite{IGO} (battery-powered mouse sensors) replaces the adoption of more powerful and
expensive complex devices (smart cameras). In order for this approach to be
effective the mentioned collective action is to guarantee that
an optimal trade-off between energy efficiency, performance,
and safety is dynamically sustained. 

We plan to express this optimal trade-off in terms of
a system-environment fit.
Obviously the formulation of the LittleSister system-environment fit
will be considerably more complex than the one introduced in the present work.
A key role will be played in particular by the LittleSister awareness organ, which 
will be used to determine the level of criticality of the current situation
and set an operative mode ranging from ``energy-saving-first''
to ``safety-first''. This operative mode 
will be included in the set of context figures
of the social behavior $\bhvrtwo{\hbox{soc}}{F}$ of LittleSister's sensors.
Depending on the requirements expressed by the current operative mode
and other context figures, the system-environment fit will vary, which will
translate in a variable selection and number of sensors to be activated.
The goal we aim to reach is being able to sustain at the same time
both maximum safety and minimum energy expenditure.

\section{Conclusions}\label{s:end}

The questions we have posed in Sect.~\ref{s:intro} have been answered, to some extent,
by defining a conceptual framework for their discussion. The nature of our framework
is behavioral and ``sits on the shoulders'' of the work carried out in the first half
of last Century by ``giants'' such as Bogdanov, Wiener,
von Bertalanffy, Boulding, and several others---in turn based on the
intriguingly modern ideas
of ``elder giants'' such as Leibniz~\cite{leibniz2006shorter} and Aristotle~\cite{Sachs}.

Within our framework we have
introduced a behavioral formulation of the concepts of supply and system-environment fit as
measures of the optimality of a system with respect to
the current conditions of the environment in which the system is deployed.

Moreover, we have suggested how complex abilities such as auto-resilience and antifragility may be
expressed in terms of behaviors able to track supply and fit measures and evolve the
systemic features in function of the hypothesized future
environmental conditions.

Practical application of the concepts in this article has been briefly discussed by
considering a strategy for optimizing the collective behavior of the mouse sensors
used in project LittleSister.

As can be clearly understood, our work is far from being exhaustive or complete. In particular discussing context figures
without referring to a ``range'', or sphere of action, makes it difficult to compare
behaviors such as auditory perception in animals.
Our future work will include extending our conceptual framework accordingly.

Another direction we intend to take is the application of our concepts towards
the design of antifragile computing systems; the Reader may refer to~\cite{DF14a}
for a few preliminary ideas about this.

\section*{Acknowledgments} 
I would like to express my gratitude to Alan Carter for helping me with
the pictures in this paper.

Many thanks to Dr. Tom Leckrone (\url{https://twitter.com/SemprePhi}) for introducing me to the work of 
Alexander A. Malinovsky (A. A. Bogdanov).

This work was partially supported by
iMinds---Interdisciplinary institute for Technology, a research institute
funded by the Flemish Government---as well as by the Flemish Government Agency for Innovation by
Science and Technology (IWT).
The iMinds LittleSister project is a project co-funded by iMinds with project support of IWT.
Partners
involved in the project are 
Universiteit Antwerpen, Universiteit Gent, Vrije Universiteit Brussel, Xetal, Christelijke Mutualiteit vzw,
Niko Projects, JF Oceans BVBA, and SBD NV.

\bibliographystyle{IEEEtran}
\IEEEtriggeratref{16}


\end{document}